\documentclass{elsart}
       \usepackage{amssymb}
       \usepackage{amsmath}

       \usepackage[xdvi]{graphicx}

\newcommand{\pam}[1]{|#1|_{p}}
\newcommand{\slaninalabel}[1]{%
\label{#1}%
}%
\def\slaninafigdir{.}
\begin{document}
\begin{frontmatter}
%



\title{How the quasispecies evolution depends on the  
topology of the genome space}


\author[MFF,FzU]{Michal Kol\'{a}\v{r}\thanksref{mkem}}
and
\author[FzU]{Franti\v{s}ek Slanina\thanksref{fsem}}

\address[MFF]{Institute of Physics of Charles University in Prague,
Ke Karlovu 5, CZ-12116, Praha, Czech Republic}
\address[FzU]{
Institute of Physics, Academy of Sciences of the Czech Republic,
Na Slovance 2, CZ-18221 Praha, Czech Republic }
\thanks[mkem]{E-mail: kolarmi@fzu.cz}
\thanks[fsem]{E-mail: slanina@fzu.cz}

\begin{abstract}
We compared the properties of the error threshold transition in
quasispecies evolution for   
three different topologies of the genome space. They are a) hypercube
b) rugged landscape modelled by an ultrametric space, and c) holey
landscape modelled by Bethe lattice. In all studied
topologies the phase transition exists. We calculated the critical
exponents in all the cases. For the critical exponent corresponding
to appropriately defined susceptibility we found super-universal
value. 
\end{abstract}

\begin{keyword}
Stochastic processes \sep biological evolution \sep ultrametric space
\PACS 05.40.-a \sep  87.23.Kg
\end{keyword}
\end{frontmatter}

\section{Introduction}\slaninalabel{introduction}

The conceptual simplicity and readiness for mathematical description,
along with obvious practical relevance makes the Darwinian
evolution a prominent subject of theoretical biology. Moreover, the
formalisation in terms of stochastic processes reveals close relations
to many problems studied before in theoretical physics. Therefore, it
is natural to look for physical tools which may answer biologists'
questions about natural evolution.  A lot of effort was devoted to
this area recently \cite{ba_fly_la_92,ba_sne_93,pa_ma_ba_96,va_au_95,fe_pla_dia_95,so_ma_96,ro_ne_96,zhang_97,drossel_98a}.

The process of the biological evolution consists in three
steps, though two of them are simultaneous: $reproduction
\rightarrowtail mutation\rightarrow selection.$ The important
thing to note is that the \emph{biological fitness} function which
denotes individuals' ability to produce viable offspring is
dependent on their phenotype. On the other hand, the mutations occur 
in the  genotype --- information
stored in a sequence of DNA.  The overwhelming complexity of the
fitness function, which assigns to each microscopic genotype
corresponding macroscopic phenotype, makes the theoretical treatment
of evolution an extremely complicated task.

Simplifications of the problem are necessary.
One of the most prominent schemes was developed within 
the \emph{quasispecies} model of the biological evolution,
introduced by Eigen \cite{eig71}. It showed to be plausible and very
useful when investigating the mechanisms of microevolution. 
In the language of physics we can roughly speak of diffusion in
a complicated potential. Originally it was introduced as
a model of chemical prebiotic evolution, but it is
also reasonable to use it for viruses and bacteria which do not haven
too complex genome. 
Among the host of other approaches, let us mention at least 
the model of coevolution described by Kauffman's NKCS
model \cite{kauffman_90a,kau91}.

Very clear and visual presentation of the evolutionary process is
to consider it as a random walk in the \emph{fitness landscape}. This
is a surface above the genome space, where the altitude of each point
is determined by the fitness attributed to the corresponding
genotype. In fact, the fitness
landscape is not static, as it strongly depends on the ever changing
 environment, which includes interactions with
(co)evolving species as well as  abiotic influences \cite{wil01}. The
movement in 
the fitness landscape is driven by the mutations and optimal
genotypes are situated on tops of peaks in the landscape.

A broad set of different fitness landscapes was used recently to
study the behaviour of the evolutionary system. These models included
both static \cite{alt01,pel01} and dynamic landscapes
\cite{wil01,nil00,pekalski_00,pek_wer_01}. They include the
\emph{sharply-peaked 
landscape} (SPL) , the \emph{Fujiyama landscape} or the
\emph{holey landscape} (HL). Several recent reviews summarise various
approaches explored
\cite{wil01,baa_gab_99,peliti_97,drossel_01}.

In the Eigen's quasispecies model an individual is characterised
by a sequence of $d$ nucleotides (or loci)
$\mathbf{s}=\{s_{1},s_{2},s_{3},\ldots,s_{d}\}$. In real situation we
should consider that there are 4 main
nucleotides A, G, C, T(U), so the alphabet contains four letters. Two
of them are pyrimidines and the 
other two are purines. To make the treatment mathematically simpler,
we will use only two-letter alphabet $\{0,1\}$. We may interpret this
simplification in two alternative ways. We can consider putting 
the digit  $0$ in the site of the sequence
when in the same position in the nucleotide acid is a pyrimidine
and $1$ if there is a purine. Or, we can consider a set
of loci and for each locus suggest exactly two possible alleles,
denoted by $0$ and $1$. 

In every position of the sequence a mutation can occur with a
probability $\mu$. In our work we consider only point mutations
(transversions and transitions or an exchange of alleles in a
certain locus) which are only a part of all processes which change
genotype and consequently the phenotype. For example we
neglect duplications which,it seems, are important for our
understanding of protein interaction networks, see
\cite{sol_pas_smi_kep_01,vaz_fla_mar_ves_01,kim_kra_kah_red_02}. The mutation is then an
exchange of $1$ and 
$0$ in a particular site in the sequence or in a certain locus.
The on site mutation rate $\mu$ is considered to be uniform for
all sites of the sequence, although this is not the case which we
see in fact. It is considered that this simplification does not
change the results of the model.

The properties of the evolutionary process, which we are going to
compare in the article, are the critical coefficients of the error
threshold transition. The \emph{error threshold} is the phase
transition found in SPL which separates two regimes of the
quasispecies evolution --- the \emph{adaptive regime} and
\emph{wandering regime}. In the adaptive regime the cloud of
quasispecies is formed around the \emph{wildtype}, whereas in the
wandering regime no wildtype is formed due to a mutation load.

The error threshold phenomenon comes out of the competition
between the selective advantage of
the wildtype $\alpha$ and the mutation load presented by the rate
$\mu$ and the genome length $d$.  The threshold is characterised
by the specific value of the selective advantage $\alpha_{c}$. From
the physical point of view, the error threshold may be thought of as
a dynamic phase transition from localised to extended ground-state
eigenvector.  
Recently it was shown \cite{kam_bor_02} that coevolution of several
species may drive the system close to the error threshold. It is
therefore highly biologically relevant to study the properties of the
error threshold transition in detail.

It is evident that the native geometry of the genome space is the
hypercube 
${\mathcal S}=\{0,1\}^d$. So, this is the natural starting point when
building a model for a fitness landscape. The main focus will be aimed
at the presence or absence of adaptive regime induced by a single
maximum in the fitness landscape. Therefore, the first thing to try 
is the sharply--peaked landscape on the hypercube. This model was solved
exactly by Galluccio et al. \cite{ga_gra_zha_96,galluccio_97}. 

However, the real landscapes are far more complicated. Indeed, it is
expected that rugged landscapes with many competing maxima represent a
realistic picture \cite{ba_fly_la_92}. Fortunately enough, such
landscapes are well--known in the theory of spin glasses
\cite{me_pa_vi_87} and a ``spin--glass'' theory of evolution was
investigated \cite{am_pe_sa_91}.

The main result coming from the spin--glass theory is the {\em
ultrametric structure} of the minima in the free--energy landscape of
spin glasses, which can be translated into ultrametric or hierarchical
structure of maxima in the corresponding fitness landscape.

Somewhat more coarse--grained description of the evolution may arise,
if we assume the ultrametric structure of fitness maxima. Indeed, we
may think of the evolution as a hopping process between various peaks,
where the system spends most of its time, while the detailed course of
the hops, composed of many individual point mutations, can be
accounted for phenomenologically by tuning appropriately the hopping
probabilities. 

Here, we will be interested again in the error--threshold transition,
now interpreted in such a way that all but one of the peaks have the
same weight, while the singled--out peak is higher and corresponds to
the ``wildtype'' of the sharply--peaked landscape. Therefore, the
treatment can be very similar to SPL from the mathematical point of
view, but differs significantly in that now we take into account, at
least approximately, the ruggedness of the real fitness landscape.

Yet another approach to the modelling of the fitness landscape is
possible, namely considering the holey landscape.
Indeed, most of the point mutations which may occur at the basic level
of the evolutionary picture are lethal for
the individual. Therefore, the hypercube does not represent a good
approximation to the evolutionary dynamics, because only few of the
hypercube's edges represent paths to possible new genomes. A sparsely
connected set of points selected at some of the hypercube corners is
perhaps better choice. 

The problem resembles the study of
topologically disordered solids, where the random network of bonds is
often well modelled by the Bethe lattice. Therefore, third geometry we
investigate in our work is that of the Bethe lattice. Again, as in the
previous cases, one of the points will be considered as the
``wildtype'' with larger fitness, while all other points will have
uniform fitness (smaller than the wildtype).

\section{General scheme for sharply--peaked and related landscapes}
\slaninalabel{sec:generalscheme}

In the quasispecies model, evolution of the size $p_{i}$ of an asexual
population for the genome represented by the lattice site $i$
is  governed  by the equation
\begin{equation}
\dot{p}_{i}(t)=\sum_{j}T_{ij}p_{j}(t)\;\; .
\slaninalabel{eq:dynamics}
\end{equation}
describing a Markov process in the genome space.

The quantity
$p_{i}$ will represent the relative population size,
or it will be proportional to the probability of a certain individual
to belong to the 
population in the site $i$. In so doing, we tacitly assume to work within
the \emph{infinite population} limit.

The time evolution may take place in discrete time; in this case the
derivative corresponds to $\dot{p}(t)\equiv p(t+1)-p(t)$. However, in
our work the continuous--time description will be used,
$\dot{p}(t)\equiv \frac{\partial p(t)}{\partial t}$, unless explicitly
stated the opposite.

The dynamical matrix can be written in the form
$T_{ij}=Q_{ij}+\delta_{ij}(A_i-D)$, where $A_{j}$ is
proportional to the fitness of the genome $i$ and so it
represents a reproductive potential of the quasispecies. The mutation
matrix  $Q_{ij}$
represents the rate of
mutation from the point $i$ in the genome space to another point $j$. The
diagonal elements of the matrix $Q_{ii}$ are the probability rates
of the perfect replication of the individual $i$. We may also
introduce the homogeneous
external stress which conserves the population size, through the
decay rate $D$.

The dynamical matrix $T_{ij}$ is supposed to be symmetric. This allows us to  
apply the common machinery of Green functions. Indeed, the main object
of interest will be the resolvent (Green function) of the dynamical
matrix
\begin{equation}
\mathcal{G}(\zeta)=\frac{1}{\zeta-T}
\end{equation}
This formalism is particularly suitable for the case of the sharply--peaked
landscape. Indeed, we suppose that the dynamical matrix can be
decomposed as $T=T^0+V$, where $T^0$ corresponds to the flat landscape
of neutral evolution and the perturbation $V$ describes the presence
of the peak localised at point $i=1$,
\begin{equation}
V_{ij}=\alpha\delta_{ij}\delta_{i1}
\slaninalabel{eq:peak}
\end{equation}

Indeed, supposing that we are able to fully solve the neutral
evolution, i. e. to find the ``free'' resolvent $\mathcal{G}^0=(\zeta-T^0)^{-1}$,
the error threshold is closely related to the existence of the
split-off pole in the resolvent $\mathcal{G}(\zeta)$. For the position
of the pole  
$z$ we have the equation (Koster--Slater condition)
\begin{equation}
\frac{1}{\alpha}=G(z)
\slaninalabel{eq:KScondition}
\end{equation}
where $G(\zeta)=\mathcal{G}^0_{11}(\zeta)$ 
is the diagonal matrix element of the free
resolvent at the site of the peak.
If the equation (\ref{eq:KScondition}) has a real solution, there is a
split-off pole, the stationary state is localised and we are in the
adaptive phase of the evolution. On the contrary, the absence of a
real solution indicates that the stationary state is extended and we
are in the wandering regime.

Our strategy in the following will consist in calculating the free
resolvent for various models of the genome space and then analysing
the solution of the equation (\ref{eq:KScondition}). To start with, we
reexamine the previously obtained results on the hypercube.

\section{Exact solution on the hypercube}
\slaninalabel{sec:hypercube}

In the article \cite{ga_gra_zha_96} Galluccio et al. investigated the
evolution of the quasispecies in the sharply--peaked landscape on the hypercubic 
lattice. The model is exactly soluble. Let us start by re-derivation of
their main results.

The following assumptions are made in \cite{ga_gra_zha_96}:

\begin{enumerate}
\item The topology of the genome space is the hypercube.

\item The evolution takes place in discrete time, i.e. $\dot{p}\equiv
p(t+1)-p(t)$. 

\item At one time step, only single point mutation occurring with
probability $\mu$ is
allowed. Therefore, the mutation matrix connects only the nearest
neighbours along the edges of the hypercube. 

\item The fitness landscape is flat with only a single point which
takes the additive selective advantage $\alpha>0$ (sharply--peaked landscape).

\end{enumerate}

The consequence of allowing single point mutations in discrete time
 evolution, is
 that they had to restrict a mean number
of mutations per genome --- genomic mutation rate $\overline\mu=\mu d$
--- to be less than 1. This need not to be valid in general, for
example in natural viral populations, as is shown in paper
\cite{dra99} where the genomic mutation rates are estimated for
measles virus, poliovirus, VSV and rhinovirus. The values are in
the range from 0.13 to 1.15 mutations per replication. In other
work \cite{sch99} the genomic mutation rates were estimated in the
range from 0.475 to 4.28 mutations per replication. This is the
reason why we decided to use a continuous time approximation. The
value $\mu$ can be easily found if we know the length of the life
cycle and the mutation rate per site per replication. We took the
rate from mentioned articles. The ``lifetime'' of measles virus or
more precisely the time required for production of a new burst of
viral particles is in the order of hours or days. So, for example,
the mutation rate in RNA virus is around $\mu= 10^{-4}$ per site
per generation \cite{dra99}. With the considered generation time
$20$ hours we get approximately $\mu=1.5 \times 10^{-9}$ per site
per second. The presented models of quasispecies have to respect
the values of $\mu$ and the length of the sequence which is in order of
kilobases. For instance, $d = 15,894$ bases for the measles virus \cite{dra99}.
Therefore, we choose the continuous-time description of the evolution
process instead. The rest of the assumptions listed above remain valid.
Let us proceed with formalisation of the above assumptions.

The genome space is the $d$-dimensional hypercube, ${\mathcal
S}=\{0,1\}^d$. The points in the genome space are the sequences of
base pairs 
$i=[i_1,i_2,...,i_d]\in {\mathcal S}$. 
Let us denote $\bar{i}(k)\in {\mathcal S}$ the sequence which
differs from $i$ by the point mutation at $k$-th base pair, 
$(\bar{i}(k))_l=i_l+(1-2 i_l)\delta_{kl}$.
Then the dynamic matrix of the neutral evolution corresponds to the
diffusion in the hypercube
\begin{equation}
T^0_{ij}=\mu\sum_{k=1}^d\delta_{j\,\bar{i}(k)}-\mu\, d\, \delta_{ij}
\end{equation}
The sharp peak located at the wildtype sequence $0\equiv [0,...0]\in
{\mathcal S}$ contributes by the term expressed by
Eq. (\ref{eq:peak}).

The general condition (\ref{eq:KScondition}) assumes very simple form,  
when we use the Fourier transform in the space ${\mathcal S}$ and 
divide the space with respect to the Hamming
distance $h$ from the wildtype. The number of sequences in each Hamming
distance class is given by the binomial coefficient ${d \choose
h}$. We obtain
\begin{equation} \slaninalabel{eq:hgalluccio}
 \frac{1}{\alpha}=\frac{1}{2^{d}}\sum_{h=0}^{d} {d \choose h}
 \frac{1}{z+2\mu h}\quad .
\end{equation}

In the limit 
$2^{d}\rightarrow\infty$ the calculation further simplifies.
The point is that the binomial
distribution of classes reduces in that limit to a very thin and
high band which we approximate by a $\delta$-function in the mean
value, ${d \choose h} \sim 2^{d}
\delta(h-d/2)$. 

However, we should be aware that this approximation holds only for $z$
not too close to $0$. On the other hand, the error threshold itself is
characterised by the fact that the solution $z$ of the equation
(\ref{eq:hgalluccio}) approaches to $0$. Therefore, we have to consider
in (\ref{eq:hgalluccio}) separately the term with $h=0$, while the
rest of the sum, for $h>0$, is approximated by the $\delta$-function
as hinted above.
Hence, the approximate form of the Eq. (\ref{eq:hgalluccio}) is\footnote{we have done numerical solutions of
(\ref{eq:hgalluccio}) and (\ref{eq:happrox}) for $d=44$ and even for such
a short sequence the approximative formula fits very well.
So, we expect very
good agreement for the genomes with the length comparable to e. g. the
measles virus's genome.} 
\begin{equation} \slaninalabel{eq:happrox}
 \frac{1}{\alpha}=\frac{1}{2^{d}}\,\frac{1}{z}+\frac{2^{d}-1}{2^{d}}\,\frac{1}{z+\mu d}.
\end{equation}
The approximation we use here goes along different line than the one
used in \cite{ga_gra_zha_96} 
but it gives the same results in lowest order. 

The solution $z$ of Eq. (\ref{eq:happrox}) is 
\begin{equation} \slaninalabel{hz}
  z=\frac{1}{2}\left(\alpha-\mu d + \sqrt{(\alpha-\mu d)^{2}+\frac{4\alpha\mu
  d}{2^{d}}}\right).
\end{equation}

Now we perform the limit
$d\rightarrow\infty$. It requires some care, as the product of $\mu d$
must stay finite, otherwise diagonal  
elements of the transfer matrix diverge. So we fix $\bar{\mu}=\mu d$
constant. Performing the limit we have to distinguish two
different cases,
\begin{eqnarray}
\quad\quad\quad\quad\quad\quad\quad\quad z=&\frac{1}{2^{d}}\,
\frac{2\alpha\bar{\mu}}{\bar{\mu}-\alpha} &\quad\mathrm{for}\quad (\alpha -\bar{\mu})<0\nonumber\\ 
\quad\quad\quad\quad\quad\quad\quad\quad z=&\alpha-\bar{\mu} &\quad\mathrm{for}\quad(\alpha-\bar{\mu})>0.
\slaninalabel{hz1}
\end{eqnarray}

As we see behaviour of the largest eigenvalue $z$ changes sharply
in the point $\alpha_{c}=\bar{\mu}$. Here we recognise the \emph{error
threshold}.
\begin{figure}
  \centering
  \includegraphics[width=0.8\columnwidth]{\slaninafigdir/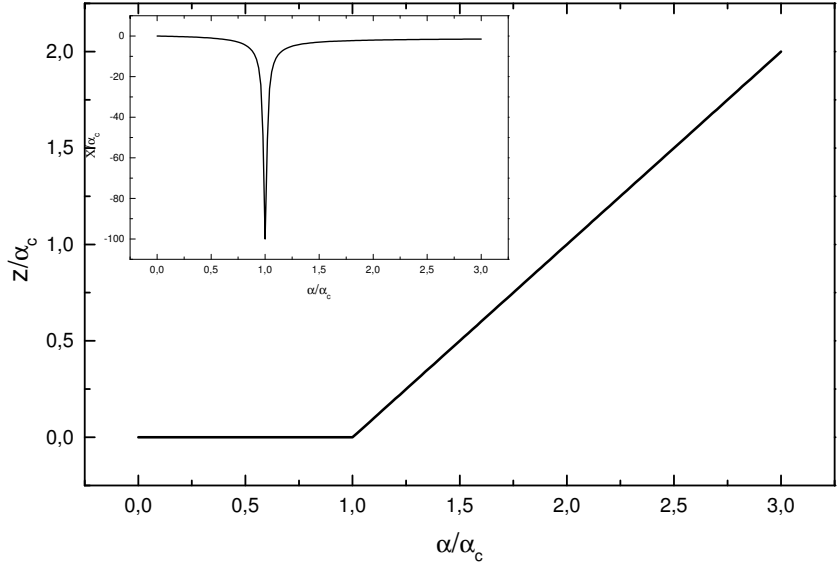}
  \caption{\small{A graphical view of $z$ and $\chi$ (inset) for the quasispecies
  evolution on the hypercubic lattice.}}
\slaninalabel{hzplot}
\end{figure}

Our next aim is to calculate the \emph{susceptibility} of the
quasispecies to a change of the sequence length. Elongation of the
sequence increases a number of possible sequences $N=2^{d}$. The
susceptibility, then, shows us how the largest eigenvalue and thus
a relative fitness of the wildtype changes with the ``volume'' of
the sequence space.

We differentiate (\ref{hz}) with the respect to $N$. Then the
limit $N\to\infty$ is done. Defining
\begin{equation}\slaninalabel{huchi}
   \chi = \lim_{N\to\infty} N^{2}\frac{\partial z}{\partial N}
\end{equation}
we get in the vicinity of the error threshold
\begin{equation}\slaninalabel{huchi2}
   \chi=-\frac{\alpha\alpha_{c}}{|\alpha-\alpha_{c}|}
\end{equation}

Now we introduce critical exponents $\beta^\pm$,
$\gamma^\pm$. They characterise
behaviour of the largest eigenvalue $z$ and the susceptibility,
respectively, close to the error
threshold,  from the left $(-)$ and right $(+)$ side.
In the
vicinity of the error threshold we assume the critical behaviour
\begin{eqnarray}
\quad\quad\alpha<\alpha_{c}\quad &\quad z\sim
N^{-1}|\alpha-\alpha_{c}|^{\beta^{-}} \quad &\quad \chi\sim
|\alpha-\alpha_{c}|^{\gamma^{-}}\quad,\nonumber\\
\quad\quad\alpha>\alpha_{c}\quad &\quad z\sim
|\alpha-\alpha_{c}|^{\beta^{+}} \quad &\quad \chi\sim
|\alpha-\alpha_{c}|^{\gamma^{+}}\quad.
\end{eqnarray}
From (\ref{hz1}) we see that $z$ behaves close to the error
threshold as a power function with the exponents $\beta^{-}=-1$
and $\beta^{+}=1$, and the Eq. (\ref{huchi2}) implies that
$\gamma^{-}=\gamma^{+}=-1.$

\section{The Quasispecies Evolution in an Ultrametric Space} 
\slaninalabel{sec:ultrametric}

The sharply--peaked landscape is certainly an oversimplified
approximation. It describes the adaptation close to a single selected
peak within the true fitness landscape. Indeed, in reality the
landscape is rugged and consists of very many concurrent peaks. In
order to investigate their combined effect, we must turn to some model
scheme.

It was realised quite a long ago that complicated free--energy landscapes
with many minima are characteristic for disordered spin systems, namely
spin glasses \cite{me_pa_vi_87}. This analogy was brought to biology within the
spin--glass model of evolution \cite{am_pe_sa_91}. The main result of
the spin--glass theory, which will be relevant for us here, is the
tree--like (or ultrametric) structure of minima in the free--energy
landscape \cite{me_pa_vi_87,ra_tou_vi_86}. We therefore assume that the same
ultrametric structure of peaks will be characteristic for the rugged
fitness landscape. Then, we will observe the evolution on a
``mesoscopic'' time--scale, where a single event will be the jump from
one peak to the other. Clearly, it consists of many ``microscopic''
events, i. e. point mutations in the DNA sequence.

The evolution
takes place in the space $\mathcal{S}$ constructed as the set of
endpoints of a regular three with $K$ levels, where at each level
every branch splits into $p$ new branches. The ultrametric tree is
illustrated in the left panel of Fig. \ref{arbitree}.

From the mathematical point of view, the ultrametric structure is
described through $p$-adic numbers and we will use the $p$-adic
Fourier transform as the main tool for the solution.  The ultrametric
space is represented by the set $\mathcal{S}=\{0,1,2,...,p^K-1\}$ endowed
with ultrametric distance between each pair of points, denoted
$|i-j|_p$ for $i,j\in\mathcal{S}$. We refer the
reader to the Appendix for necessary definitions and formulae.
The diffusion in the ultrametric space was already thoroughly studied
before in the context of spin-glass dynamics
\cite{pa_ste_ab_an_84,og_ste_85,pa_me_dedo_85,ho_si_88,ap_fe_98,bou_dea_95}
and the concept of $p$-adic Fourier transform was developed quite long
ago in the mathematic literature \cite{taibleson_75}, and was used
implicitly (see e. g. \cite{me_pa_91a}) but its use in physics was
made explicit only recently
\cite{ded_ca_te_97,pa_so_99,av_bi_ko_99,av_bi_ko_os_02}.

The ultrametric structure will be reflected by the fact that the
probability of jumps (mutations) from one point to the other will
depend only on the ultrametric distance between the two points. 
It is analogous to the hypercube, where the mutation probability
depends only on the Hamming distance.

\begin{figure}
  \centering
  \includegraphics[width=0.45\columnwidth]{\slaninafigdir/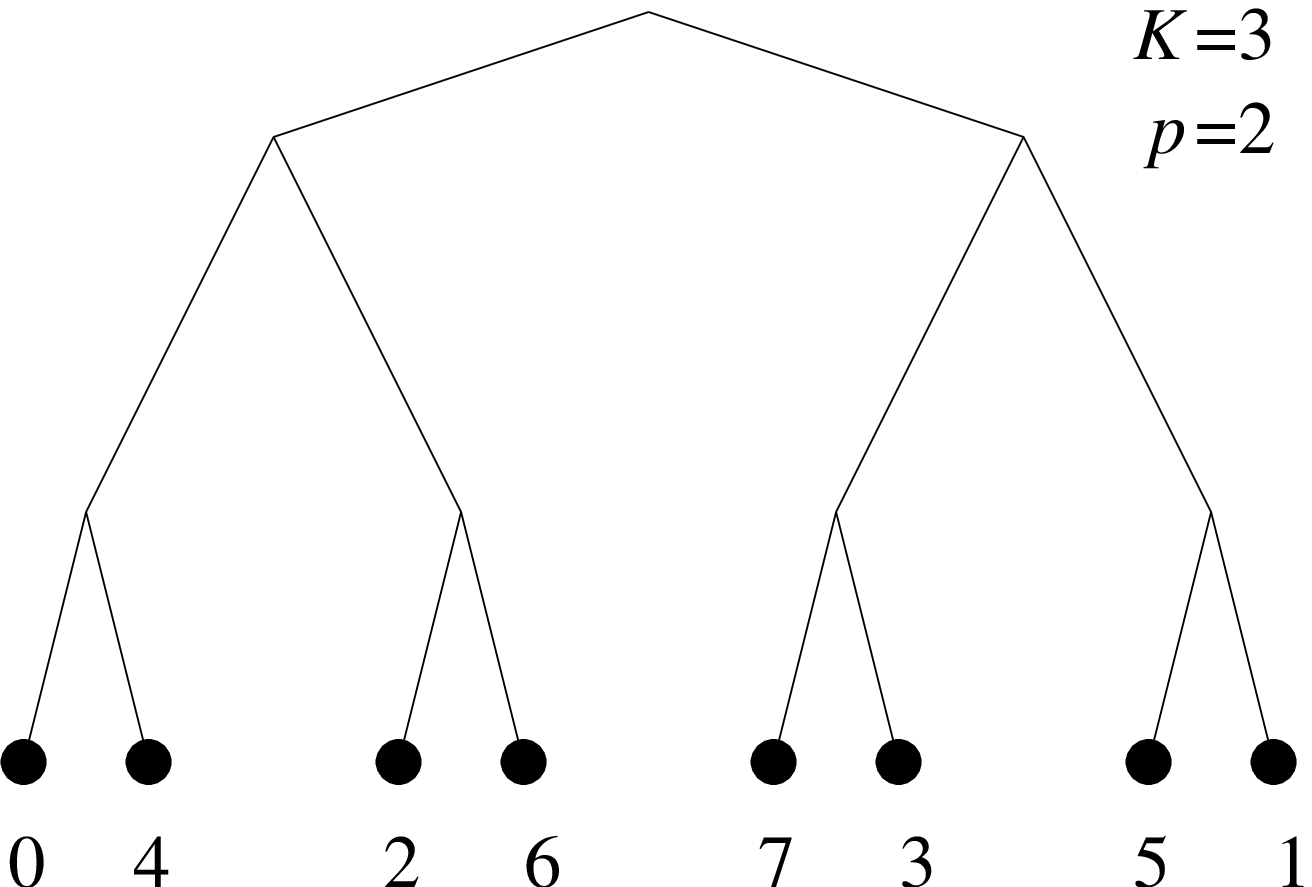}
\hspace*{5mm}
  \includegraphics[bb=0 0 530 390,width=0.45\columnwidth]{\slaninafigdir/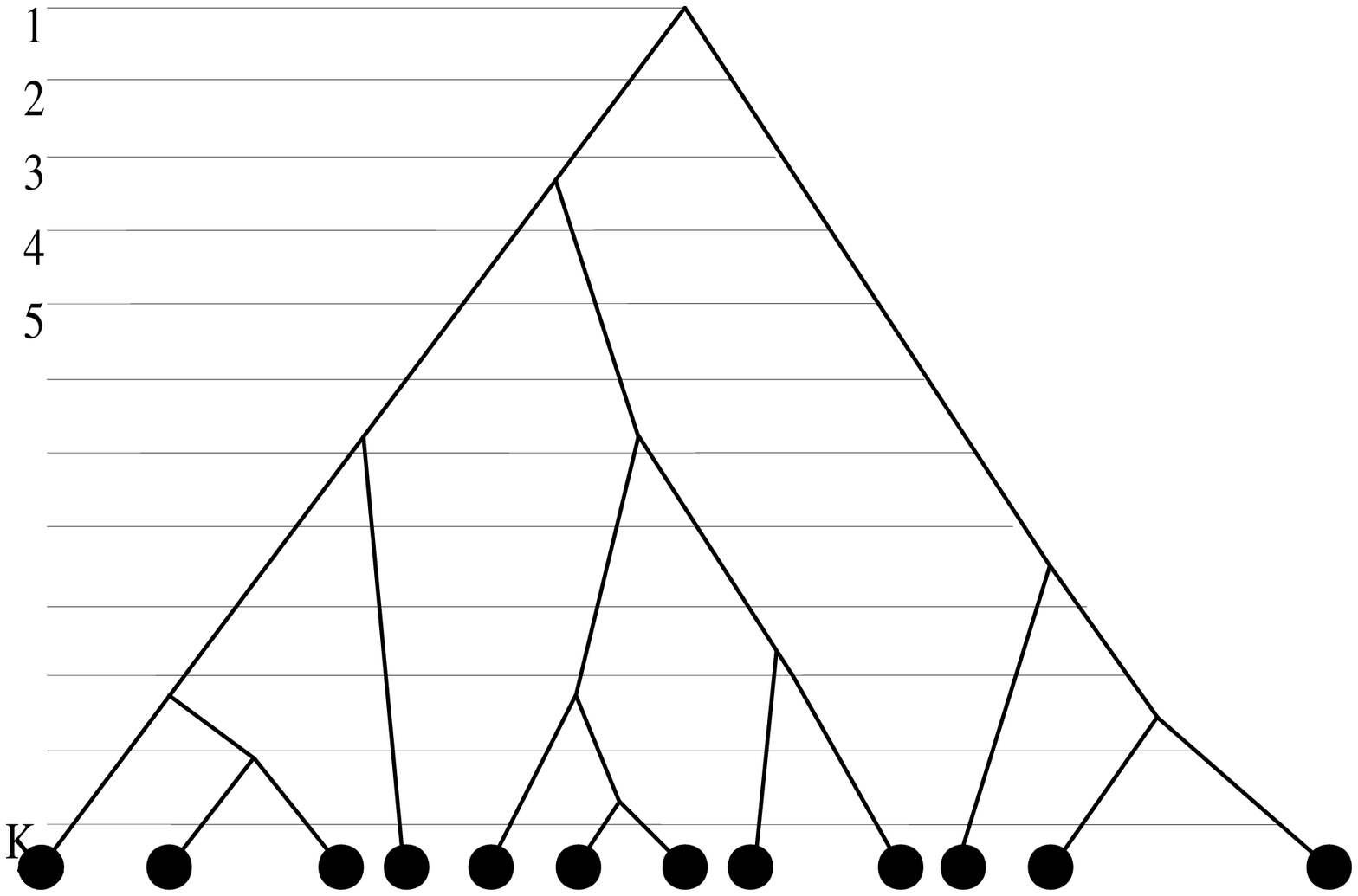}
  \caption{(Left) The ultrametric tree with three levels and with the
  branching $2$. Below are written $2$-adic numbers. (Right) Division of an arbitrary
  tree in $K$ levels with an imperfect branching $p\rightarrow 1^{+}$. The difference $p-1$ is
  proportional to the probability of branching in an arbitrary level.}
\slaninalabel{arbitree}
\end{figure}

Movement of the individual on the ultrametric lattice is driven by
the dynamic matrix $T$ which is similar to the one used on
the hypercubic lattice. There is one important difference --- in the
previous case jumps only to the nearest neighbours were
considered, but now we must allow a possibility to jump further,
otherwise we get evolution only in one isolated branch of the
tree.

The evolution in ultrametric analog of the sharply--peaked landscape 
is described again by the equation  (\ref{eq:dynamics})
where now the
``unperturbed'' matrix $T^0$ reflects the symmetry of the ultrametric
space.
This means that the value of the matrix element  $T^0_{ij}$ depends only
on the ultrametric distance between the points $i$ and $j$.  The
``perturbation''  $V$ corresponds to the single site with selective advantage.
We can write explicitly
\begin{equation}
\begin{array}{lll}
T^0_{ij}=&\,\tilde{T}&\quad\text{if}\quad i=j\\
T^0_{ij}=&\,T_{m}&\quad\text{if}\quad \pam{i-j}=p^{-m}\quad,  \\ 
V_{ij}=&\,\alpha\,\delta_{ij}\delta_{i0}&\quad.
\end{array}
\end{equation}

Moreover, we will require that the bare transition matrix conserves
probability. This fixes the value of the diagonal element, 
\begin{equation}
\tilde{T}=-\sum_{b (\ne a)}T_{ab}=-\sum_{m=0}^{K-1} (p-1)p^m\,T_m
\slaninalabel{eq:diagonal-ultra}
\end{equation}
This requirement is not substantial for our treatment, but simplifies
somewhat the formulae. Especially it has the consequence that the
matrix $T^0$ has an uniform eigenvector $(1,1,...,1)$ with corresponding
eigenvalue equal to 0.

Thus, the fitness landscape is fully described by the sequence of $K$
numbers $T_0, T_1, ..., T_{K-1}$ and the parameter $\alpha$. As
$T^0_m$ is the probability of hopping to the distance $p^{-m}$, it is
natural to require that 
\begin{equation}
T^0_{m+1}>T^0_m
\slaninalabel{eq:Tisgrowingsequence}
\end{equation}

To follow the general scheme of section \ref{sec:generalscheme} we
should first find the unperturbed 
resolvent $\mathcal{G}^0(\zeta)=(\zeta-T^0)^{-1}$. So, we need to diagonalise the
matrix $T^0$.

To this end we use the $p$-adic Fourier transform. 
For details we refer the reader to the Appendix. The basic result is,
that the eigenvalues of the matrix $T^0$ are $\hat{T}_m$,
$m=0,1,...,K$, where
\begin{equation}
\begin{split}
\hat{T}_{0}=&\,\tilde{T} + \sum_{l=0}^{K-1}
T_{l}p^{K-l-1}(p-1)\\ 
\hat{T}_{m}=&\,\tilde{T} +
\sum_{l=m}^{K-1} T_{l}p^{K-l-1}(p-1)-T_{m-1}p^{K-m}, \quad
m=1,\ldots,K.
\end{split}
\end{equation}
From the conditions (\ref{eq:diagonal-ultra}) and
(\ref{eq:Tisgrowingsequence}) we can immediately deduce that 
 $\hat{T}_{0}=0$ is the maximum eigenvalue, all other eigenvalues are
negative and $\hat{T}^0_{m+1}<\hat{T}^0_m$. Using the inverse $p$-adic
Fourier transform we get for the unperturbed resolvent the expression
\begin{equation}
G(z)=\frac{1}{p^{K}}\big[ \frac{1}{z} + \sum_{l=1}^{K}
\frac{p^{l-1}(p-1)}{z-\hat{T}_{l}}\big]\quad .
\end{equation}
Introducing the normalised density of states (DOS) as
\begin{equation}
\mathcal{D}(\hat{T})=\frac{1}{N}(\delta(\hat{T})+\sum_{j=1}^{K}
p^{j-1}(p-1)\delta(\hat{T}-\hat{T}_{j}))
\slaninalabel{eq:DOS}
\end{equation}
we can write the Koster-Slater condition (\ref{eq:KScondition}) in general form
\begin{equation}
\frac{1}{\alpha}=\int\frac{\mathcal{D}(\hat{T})}{z-\hat{T}}\quad .
\slaninalabel{eq:KSwithDOS}
\end{equation}
Clearly, the factor $p^{j-1}(p-1)$ is the multiplicity of the
eigenvalue $T_j$.

Let us come back to biological aspects. Each possible quasispecies has
got a certain number of alleles, (or sets of subsequences) but we
can not consider that every combination of those alleles forms a
quasispecies which is able to survive. 
In fact, the ultrametric tree need not be regular but bears some level
of randomness. The ultrametric space considered so far was regular,
characterised by fixed branching ration $p$. We shall take into
account the randomness by the following approximate procedure.

Let us take an arbitrary tree, maybe random.
With the number of distinct end-points $N=p^K$ kept fixed we enlarge the
number of levels, $K\to\infty$. We can freely do so, as the
definition of levels is arbitrary: if no branches appear at certain
level, we simply have $p=1$ locally. With this definition, $p$ may
assume different values at different levels as well as at different
branches within the same level. The randomness of the tree is then
translated into the set of random $p$-s. We illustrate the situation
in the Fig.~\ref{arbitree}.  The approximation consists in
characterising the tree by an average value of $p$, which now may be
any real number larger than $1$. Furthermore, we will assume that the
properties of trees with non-integer values of $p$ can be
approximately calculated by analytical continuation from the results
obtained for regular, non-random trees and arbitrary integer $p$.
Moreover, we should keep in mind that with $N$ fixed and diverging
number of levels $K\to\infty$ we should make the limit $p\to 1^+$.
At the end of the calculations, we will make also the thermodynamic
limit $N\to\infty$.

Therefore, we take the formulae obtained before for general $p$
and suppose the variable $p$ is now a real number. Then, we will
perform the limits
\begin{equation}
\begin{split}
\text{first:}\quad&
K\to\infty,\; p\to 1^+\quad \text{with}\;N=p^K\;\text{fixed}\\
\text{next:}\quad&
N\to\infty
\end{split}
\slaninalabel{eq:limitprocedure}
\end{equation}

Now we should specify the form of the jump rates  $T_{m}$.
The rate $T_{m}$ of a certain
individual to mutate to some other with the sequence which is in
the ultrametric distance $p^{-m}$ from the original one will be
considered in the form 
\begin{equation}
T_{m}=\mu\,p^{(m-K)\tau}\quad .
\end{equation}
 We solve
this case for the value of $2>\tau>1$. We will see later that this is
indeed the interval relevant for the error-threshold phenomenon.

The choice of an arbitrary power $\tau$ allows simulation of a
quite general situation. In the limit $p\rightarrow 1$ possible
mutation rates to distinct sites are in a very wide range,
starting in the value $\mu p^{-(K-1)\tau}$ and ending with the
highest possible jump rate $\mu$. And this is exactly, in our
opinion, what we find in reality --- a probability rate of the
jump from one codon sequence (or a certain set of alleles) to
other is not uniform.

In order to solve the equation (\ref{eq:KSwithDOS}) we need to
compute the density of states (\ref{eq:DOS}).
The
eigenvalues of $T$ can be easily found using the Fourier
transform. A value of the diagonal element $\tilde{T}$ is again
established with the help of the condition that the sum of
elements of the matrix's column equals zero, i. e. $\hat{T}_{0}=0$.
\begin{equation}
\hat{T}_{m}=
-\frac{\mu}{p}\,\frac{p-1}{p^{\tau-1}-1}\,p^{-(\tau-1)\,K}
\left(
  p^{(\tau-1)\,m}
  \left(
    1+\frac{p^{\tau-1}-1}{p^{\tau-1}(p-1)}
  \right)
  -1
\right)
\slaninalabel{eq:FTofT}
\end{equation}

The density of states
$\mathcal{D}(\hat{T})$ will be calculated in the limit (\ref{eq:limitprocedure}).  Its support can be found easily by computing
the lowest eigenvalue $\hat{T}_K$, which in the limit
(\ref{eq:limitprocedure}) is equal to
$\hat{T}_\infty=-\frac{\mu\,\tau}{\tau-1}$. 
 Substituting
$x=p^{m}$ into (\ref{eq:FTofT})
and inverting the dependence we obtain the function $x(\hat{T})$, hence
\begin{equation}
\mathcal{D}(\hat{T})
=
\frac{1}{N}\,\left|\frac{\mathrm{d}x}{\mathrm{d}\hat{T}}\right|
=
\frac{1}{\tau-1}
\left(
  \frac{\tau-1}{\mu\,\tau}
\right)^\frac{1}{\tau-1}
\left(
  -\hat{T}
\right)^\frac{2-\tau}{\tau-1}
\end{equation}
for $\hat{T}\in (-\frac{\mu\,\tau}{\tau-1},0)$ and
$\mathcal{D}(\hat{T})=0$ outside this interval. 

The equation for the split-off eigenvalue $z$ now becomes
\begin{equation}
\frac{(\tau-1)\,\bar{\mu}^{\,\frac{1}{\tau-1}}}{\alpha}
=
\int_0^{\bar{\mu}}
\frac{y^\varepsilon}{z+y}
\mathrm{d}y
\slaninalabel{eq:forz0ultratau}
\end{equation}
where we denoted 
\begin{equation}
\varepsilon=\frac{2-\tau}{\tau-1}\quad .
\end{equation}
and
\begin{equation}
\bar{\mu}=\frac{\mu\,\tau}{\tau-1}
\end{equation}
The integral on the RHS of (\ref{eq:forz0ultratau}) is, for
non-integer $\varepsilon$,
\begin{equation}
\int_0^{\bar{\mu}}
\frac{y^\varepsilon}{z+y}
\mathrm{d}y
=
\frac{1}{\varepsilon}\,\bar{\mu}^\varepsilon
-\frac{\pi}{\sin\varepsilon\pi}\,{z}^\varepsilon+{z}\,\bar{\mu}^{\varepsilon-1}\,\sum_{k=0}^\infty
\frac{1}{k+1-\varepsilon}\left(-\frac{z}{\bar{\mu}}\right)^k\quad .
\slaninalabel{eq:intonrhs}
\end{equation}
(For integer values of $\varepsilon$ the infinite series becomes a finite
sum accompanied by a logarithmic correction).
When analysing the equation (\ref{eq:forz0ultratau})
we must distinguish two regimes. For $0<\varepsilon<1$
the leading contribution in (\ref{eq:intonrhs}) is of order
${z}^\varepsilon$, while
for $\varepsilon \ge 1$ the leading contribution is
linear in $z$. The former regime applies for $2>\tau>\frac{3}{2}$
while the latter holds for $1<\tau<\frac{3}{2}$.
This gives for the critical exponent at the error threshold
\begin{equation}
\beta^+ = \max(1,\frac{\tau-1}{2-\tau})\quad .
\end{equation}
The location of the error threshold is determined by the first term on
the RHS of the Eq. \ref{eq:intonrhs}
\begin{equation}
\alpha_c=\mu\tau\,\frac{2-\tau}{\tau-1}
\end{equation}
We can see that the error threshold vanishes in the limit $\tau\to 2$ and
diverges for $\tau\to 1$ which justifies the choice of $\tau$ within the 
interval
$(1,2)$. We investigated also the case $\tau=1$, where we send
$\mu\to 0$ simultaneously with $N\to\infty$ with $\tilde{\mu}=\mu\ln
N$ kept fixed. In this case we also observe the error threshold for
certain value of $\tilde{\mu}$, with exponent $\beta^+=1$.

\section{Evolution on Bethe lattice}
\slaninalabel{sec:bethe}

Another approach which takes into account the complicated structure of
real fitness landscapes uses the so-called holey landscape. This
concept reflects the fact that most of the random point mutations
which may occur are lethal and the evolution ends in a dead end
there. So, we may start with the hypercubic lattice $\{0,1\}^d$ as in
section  \ref{sec:hypercube} but leave only viable sites, while all
lethal genomes are deleted. The remaining lattice has highly random
geometry and the symmetry properties of the hypercube, which enabled
us to use the Fourier transform, are no more applicable here.  Thus,
another scheme of the solution should be looked for.

In fact, topologically disordered lattices are widely studied in
solid-state theory.
The Bethe lattice is a well-known model of random lattices. It has
got that large advantage that many problems
can be solved analytically. It is widely used in
statistical physics, see \cite{baxter_82}.

We take the infinite Bethe lattice with the connectivity $k$. In
each site there is a population of individuals. With the uniform
probability rate $\mu$ an individual may jump (mutate) to one of
$k$ nearest neighbours' site. Again one site with a selective
advantage $\alpha$ is present and we will study creation of the
quasispecies around that site.
The unperturbed dynamical matrix $T^0$ corresponds to the topology of the
Bethe lattice. The off-diagonal terms will be $T^0_{ij}=\mu$ if $j$
and $i$ are neighbours on the lattice and $T^0_{ij}=0$ elsewhere. The
diagonal terms on a lattice with connectivity $k$ are obviously
$T^0_{ii}=-k\,\mu$ for all $i$.

Following the general scheme of Sec. \ref{sec:generalscheme} we
calculate the diagonal element of the resolvent
$\mathcal{G}^0=(\zeta-T^0)^{-1}$ at the selected
site. As in the infinite Bethe lattice all sites are equivalent, 
we can calculate the diagonal element at any site inside the Bethe
lattice. This will be done using the partitioning (projector) method.

The procedure is based in splitting the Bethe lattice into separate
parts. This will be done in
two steps, as illustrated in Fig.~\ref{matrices}. We calculate the
diagonal element $G(\zeta)=\mathcal{G}^0_{11}(\zeta)$ at site $1$.
First, we separate  the central site $1$ from the rest of the
lattice. Because there are no loops, the rest of the Bethe lattice 
now consists of $k$ disconnected and identical infinite branches. 
The branches start at \emph{terminal} sites denoted by $2$.

\begin{figure}
  \centering
  \includegraphics[bb=0 0 340 380,width=0.5\columnwidth]{\slaninafigdir/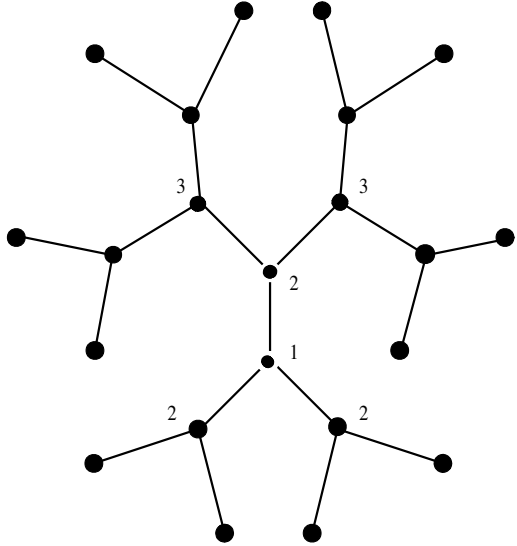}\hspace{0.5cm}
  \caption{The Bethe lattice with the connectivity $k=3$
  and the separation of branches done during the partitioning.
  The first partitioning takes part in the site $1$,
  the second one in one of the sites $2$.}
\slaninalabel{matrices}
\end{figure}

Let $P$ be the projector to the site $1$ and $Q=1-P$ the complementary
projector, which separates the rest of the lattice.
Then, we get for the diagonal element of the resolvent
\begin{equation}
G(\zeta)=P\mathcal{G}(\zeta)P=\frac{P}{\zeta-PT^0P-\Sigma(\zeta)}\quad ,
\slaninalabel{bg}
\end{equation}
where the self-energy has the form
\begin{equation}
\Sigma(\zeta)=PT^0Q\frac{Q}{\zeta-T^0}QT^0P\quad .
\end{equation}
By extracting the site $1$ the lattice consists of $k$ identical
parts. So, $\Sigma(\zeta)$ is a sum of $k$ identical contribution,
each of them coming from one isolated branch. We can write
\begin{equation}
\Sigma(\zeta)=\mu^{2}k\Gamma(\zeta)\quad ,
\end{equation}
where $\Gamma(\zeta)$ is the matrix element of the projected resolvent
$(\zeta-QT^0Q)^{-1}$ at any of the  terminal sites denoted by $2$ in
Fig.~\ref{matrices}.

The next step of the partitioning takes one branch and separates the
terminal point $2$ from it. As a result, $k-1$ disconnected branches
arise. Their terminal points are denoted as $3$ in Fig.~\ref{matrices}.
An equation analogous to (\ref{bg}) is found also for $\Gamma(\zeta)$.

We may iterate the same procedure arbitrary number of times. But if we
consider infinite Bethe lattice, the branch starting with site $2$
must be identical to any of the branches starting at site $3$. This
leads to closure in the relation for the quantity $\Gamma(\zeta)$. 
\begin{equation}
\Gamma(\zeta)=\frac{1}{\zeta+\mu k-\mu^{2}(k-1)\Gamma(\zeta)}\quad .
\slaninalabel{eq:forGamma}
\end{equation}

Solving the equations (\ref{eq:forGamma}) and (\ref{bg}) we find
the function $G(\zeta)$. From the two roots of the quadratic equation
the proper one is chosen by the requirement that
the resolvent must have correct asymptotic behaviour $G(\zeta)\sim 1/\zeta$
for $|\zeta|\to\infty$. 

However, one complication arises here. If we calculate the density of states
$\mathcal{D}(\zeta)=\frac{1}{\pi}\lim_{\varepsilon\to
0}\text{Im}\,G(\zeta-\text{i}\varepsilon)$, we observe for $k>2$ a gap arising between the
upper edge of the density of states and the point $\zeta=0$. This
reflects the well-known pathology of the Bethe lattice, which stems
from the fact that the surface points of the Bethe
lattice constitute finite fraction of the whole even in the
thermodynamic limit, unlike lattices embedded in $d$-dimensional
Euclidean space, where the fraction of the surface vanishes as
$N^{-1/d}$ when the number of sites $N$ goes to infinity. When
investigating the diffusion on the Bethe lattice, it has the unnatural
consequence that the probability density flows out constantly toward the 
surface and tends to zero in any finite portion of the lattice. In
order to compensate for this unnatural effect, we put uniform
probability flux into the lattice. In the context of biological
evolution, this amounts simply to addition of a constant to the
fitness of all genomes. Such an uniform change dos not affect the
natural selection and evolutionary competition, but simplifies the
mathematical treatment.

Indeed, introducing the uniform probability influx 
amounts simply to shifting the variable 
$\zeta\rightarrow \zeta+\mu k-2\mu\sqrt{k-1}$  so that the upper edge
of the density of states is located at $\zeta=0$. The shifted
resolvent is then
\begin{equation}
G(\zeta)=\frac{2(k-1)}{(k-2)(\zeta+2\mu\sqrt{k-1})+k\sqrt{\zeta^{2}+4\mu
\zeta\sqrt{k-1}}}\quad.
\slaninalabel{eq:Gforbethe}
\end{equation}

The real and imaginary part of the shifted
resolvent are shown in Fig.~\ref{bresolventplot}.
\begin{figure}
  \centering
  \includegraphics[width=0.7\columnwidth]{\slaninafigdir/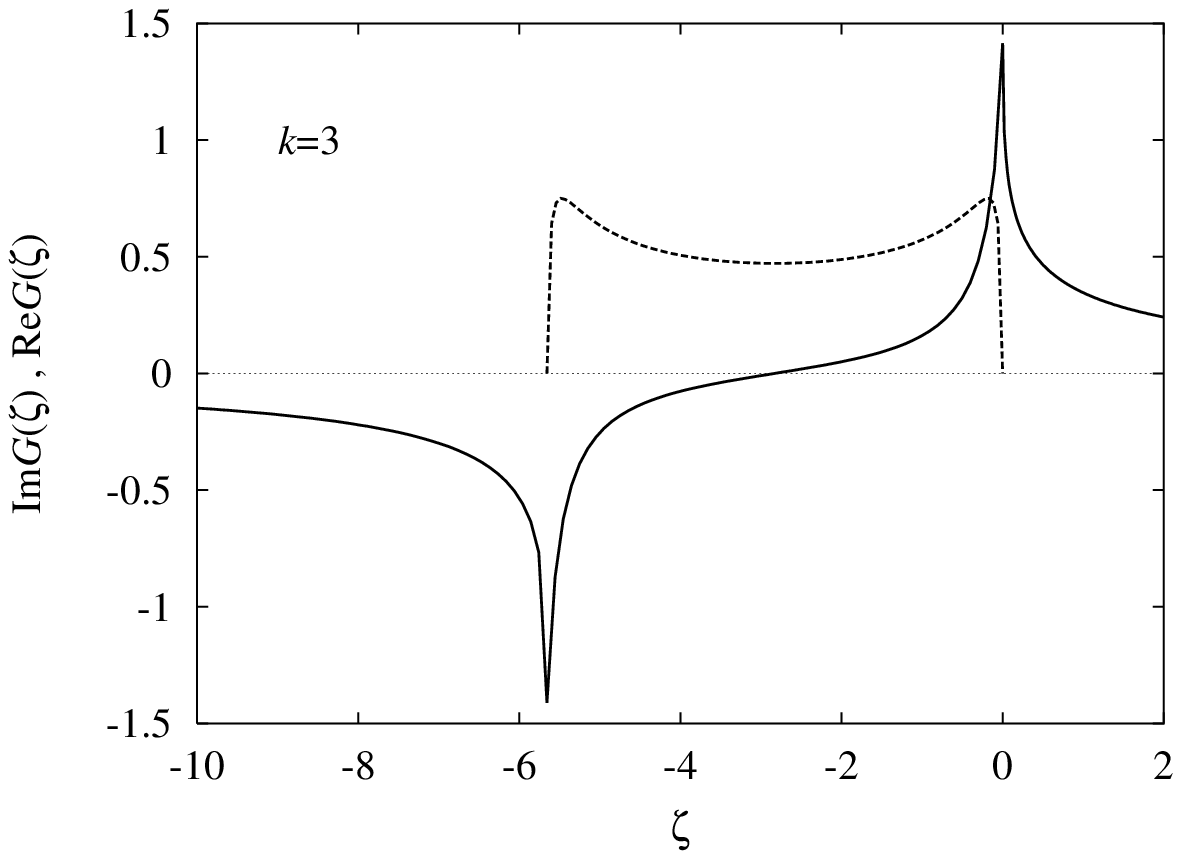}
  \includegraphics[width=0.7\columnwidth]{\slaninafigdir/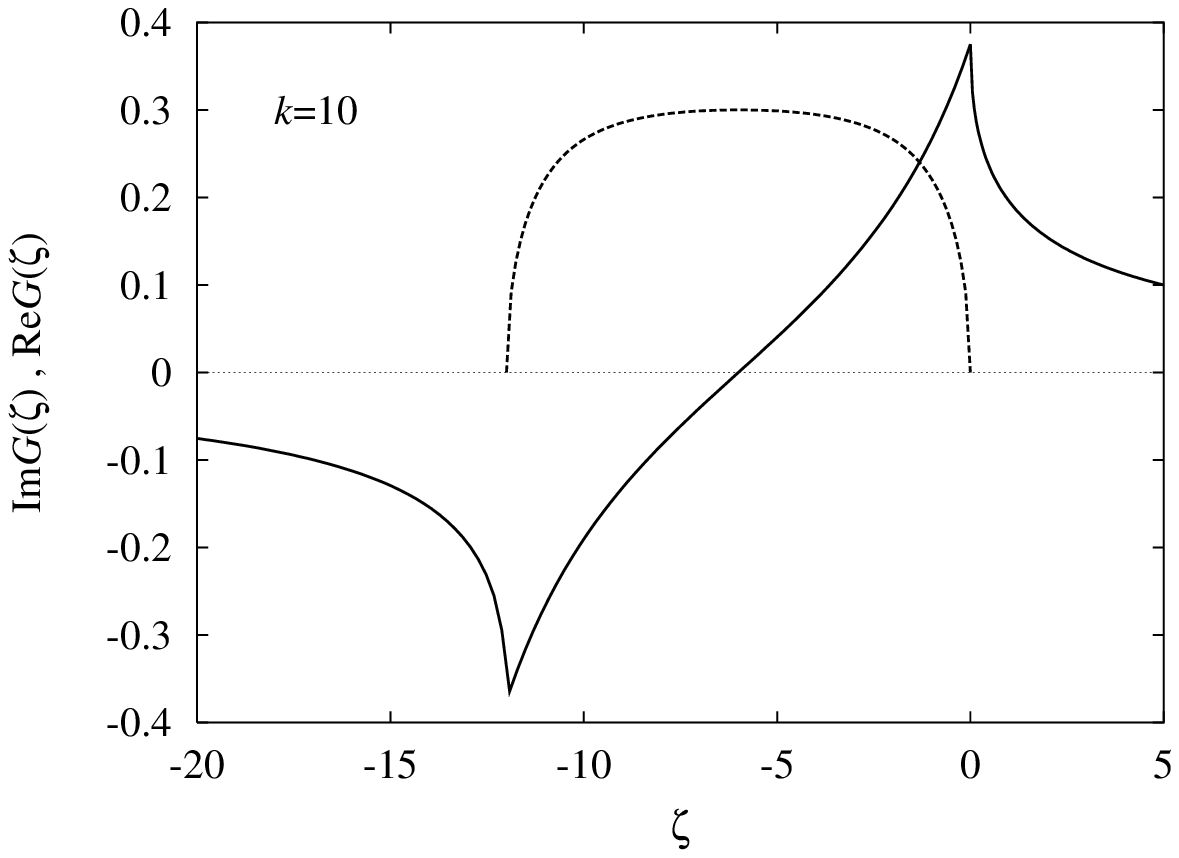}
  \caption{Real (full line) and imaginary (dashed line) part of the shifted resolvents for $k=3$ and $k=10$.}
\slaninalabel{bresolventplot}
\end{figure}

Now we can turn to the solution of (\ref{eq:KScondition}) using the
expression (\ref{eq:Gforbethe}). Since we are
looking for the real eigenvalue $z\geq 0$ we obtain
\begin{equation} z=\frac{1}{2}\left[(2-k)\alpha
- 4\sqrt{k-1}\mu + k\sqrt{\alpha^{2}+4\mu^{2}}\right].
\end{equation}
This solution is valid only above the error threshold: $\alpha\geq
\alpha_{c}=\mu(k-2)/\sqrt{k-1}$.
For smaller $\alpha$ the equation
(\ref{eq:KScondition}) have no real solution, there is no
localised state and  
the largest
eigenvalue of the dynamic matrix is $z=0$.

So we have again found that there is for $k>2$ the error threshold
between the random wandering regime and the adaptive regime at some
positive value $\alpha=\alpha_c$. For
$k=2$ the error threshold disappears, $\alpha_{c}=0$. 
However, this case is of little interest for us now, because the Bethe
lattice reduces to a linear chain, hardly being a good approximation for
the holey fitness landscape we started with.

Close to the the critical
point, we obtain that for $k>2$ the largest eigenvalue $z$ behaves
as a second power of the distance from the error threshold
\begin{equation}
z\simeq\frac{\sqrt{k-1}}{4\mu}(\alpha-\alpha_{c})^{2}\mathrm{,\quad
when}\quad \alpha\rightarrow\alpha_{c}^{+}.
\end{equation}
So we have got the critical exponent $\beta_{b}^{+}=2$. 
This reflects the behaviour of the density of states at the band
edge, which is $\mathcal{D}(\zeta)\sim |\zeta|^{1/2}$ for any $k>2$.
In fact, we obtained the relation between the exponent $\beta^+$ and
the behaviour of the density of states at the band edge already in
Sec. \ref{sec:ultrametric} when we investigated the evolution in
ultrametric space. 

\section{Super--Universality of the Critical Exponent $\gamma^{+}$}

In all previous cases the critical exponent $\beta^{+}>0$ has been
found. In this section we show that if we consider power behaviour
of the largest eigenvalue $z$ near the critical point
$\alpha_{c}$, we can find the value of the critical exponent
$\gamma^{+}$.

Assume that the whole sequence space contains $N$ distinct points.
Then, there are exactly $N$ eigenvalues of the matrix $T$.
It is known that the largest eigenvalue $0$ of the unperturbed
dynamic matrix $T^0$ is non-degenerate. 
If $N$ is very large, we can approximately write for the
matrix element of the resolvent $G(\zeta)\simeq \overline{G}(\zeta)$
corresponding to the diffusion on the lattice with $N$ sites that
\begin{equation}
\overline{G}(\zeta)=\frac{N-1}{N}G(\zeta)+
\frac{1}{N}\,\frac{1}{\zeta}\quad.
\end{equation} 
We
want to know what is the dependence  of the split-off pole $z$ on the
genome size $N$. It 
is described by the susceptibility defined in
(\ref{huchi}).

The Koster-Slater condition written as
\begin{equation}
\overline{G}(z)=\frac{1}{\alpha},
\end{equation}
which corresponds to
(\ref{eq:KScondition}), can be regarded
as an implicit function $z(N)$, so we use the formula for the
derivative of the implicit function in order to get the susceptibility
$\chi$ defined in (\ref{huchi}).
In the thermodynamic limit $N\rightarrow\infty$ the expression
simplifies and we find that the
susceptibility of the quasispecies to the change of the sequence
length is
\begin{equation}
\chi=-\frac{G(z)-\frac{1}{z}}{\frac{\partial G}{\partial z}}\quad .
\end{equation}

As a next step, in order to find $\gamma^{+}$, we suppose that in
the vicinity of the error threshold the highest eigenvalue can be
replaced by the term $z\simeq q (\alpha-\alpha_{c})^{\beta^{+}}$, where
$q$ is an appropriate coefficient. So, close to the error threshold 
\begin{equation}
G(z)\simeq \frac{1}{\alpha_{c}}-
\frac{1}{\alpha_{c}^2}\left(\frac{z}{q}\right)^{1/\beta^{+}} 
\end{equation}
hence
\begin{equation}
\chi=-\frac{\beta\alpha_c^{2}}{\alpha-\alpha_{c}},\quad\text{for}\quad \alpha\to\alpha_c
\end{equation}
which corresponds exactly to (\ref{huchi2}). The value of the critical
exponent
$\gamma^{+}=-1$ follows, and we see that the exponent does not depend
on the model in question. This proves the super-universality of the
exponent $\gamma^+$.

\section{Conclusions}

We investigated the transition between adaptive and neutral phase in
biological evolution,
which occurs at the error--threshold value of the mutation rate. We
suggested to take into account the complicated structure of the 
fitness landscape by effectively choosing non-trivial topology of the
genome space. In all cases we observed the formation of a quasispecies
around a site with selective advantage.

We compared the evolutionary process in three different
topologies. The first one was the hypercube.  (We re-derived by a
different method the already known results and completed the
study by calculating the susceptibility, not studied before.)
This corresponds to standard sharply--peaked landscape, with completely flat
landscape as a background. Second, inspired by the theory of spin
glasses, we modelled the ruggedness of the
fitness landscape by taking as the basic space the set of peaks
hierarchically organized in an ultrametric space. Third, we took the
holey fitness landscape, in which all viable genomes have equal
fitness and all lethal genomes are cut off. We approximated it by the
Bethe lattice, as usual in various problems dealing with randomly
connected networks in high dimensions.

The problem was treated as diffusion in the underlying space, the site
with selective advantage acting as a source. The formation of a
localized state (quasispecies) around the selected site was studied by
observing the behavior of the maximum eigenvalue of the dynamical
matrix governing the diffusion. If the maximum eigenvalue is split off
from the rest of eigenvalues (the band), the state is localized and we
are in the adaptive phase.

The split-off sets on at the critical value of the selective
advantage, corresponding to the error threshold.
We calculated the value of the critical value in all three
topologies. We therefore confirmed that the very existence of the
error threshold phenomenon does not depend on what is the structure of
the genome space, at least what concerns the types of topology studied
here.

As a measure of the critical behavior we chose the approach of the
maximum eigenvalue to zero when the selective advantage approaches its
critical value. We found the behavior be dominated by a power and
calculated the critical exponent. We found that it is always larger or
equal to $1$. For the hypercube it is $1$, for the Bethe lattice it has
value $2$, while in the ultrametric space it may assume any value larger
or equal to $1$, depending on details of the geometry.

In order to find the dependence between the largest eigenvalue
and the volume of the genome space, we calculated the
susceptibility of the quasispecies to the change of the sequence
length. The susceptibility diverge in the vicinity of the error
threshold. This could be of the partial interest for biologists,
since natural viral population seems to have mutation rate very
close to the critical point, see \cite{hol90}.
We found that the divergence of the susceptibility at the error
threshold is governed by a power-law with super-universal,
model-independent exponent $-1$.

All the topologies which we considered were regular. Our further
outlook is to study the influence of the randomness in the topology of
the genome space to the
evolutionary process.

\section*{Acknowledgement}
Authors would like to thank to Jan Ma\v{s}ek for critical reading
of the manuscript. Michal Kol\'{a}\v{r} would like to thank to
Vladislav \v{C}\'{a}pek for fruitful discussions and to Anton
Marko\v{s} for helpful discussions on the topic of biological
evolution, too. 

\section*{Appendix}

We will present the basic facts on ultrametric spaces, $p$-adic
numbers and $p$-adic 
Fourier transform here \cite{ra_tou_vi_86,taibleson_75,ded_ca_te_97,pa_so_99}.

The ultrametric space is a special kind of a metric space. It is a set
$\mathcal{U}$ of points endowed by a non-negative function expressing
the distance 
between two points in the space. For reasons which will
become clear later, in this article we will use the notation $|a-b|_p$
for the distance between points $a,b\in\mathcal{U}$.

As usual in metric spaces, we require that the distance from a point
to itself is zero and distance to any other point is positive.
The most important property of ultrametric spaces is that
instead of the triangle inequality a stronger condition holds,
\begin{equation}
  \pam{a-b} \leq \max\{\pam{a-c},\pam{c-b}\}\quad\forall a,b,c\in\mathcal{U}.
\end{equation}
This implies  that any triangle is equilateral or
isosceles with a small base, that every point inside a ball is its
centre, or that the diameter equals to the radius. More detailed
presentation can be found in
\cite{ra_tou_vi_86}.

There is a class of the ultrametric spaces formed by \emph{p-adic}
numbers. These are defined as sums
\begin{equation}
  a=\sum_{i=0}^{K-1} a_{i}p^{i},
\end{equation}
where $p$ and $a_{i}$, $0\leq a_{i}\leq p-1$, are natural
integers. Clearly $a\in\{0,1,...,p^K-1\}$. The distance $\pam{\cdot}$
is defined as 
$\pam{a-b}=p^{-j}$ where $p^{+j}$ is the largest divisor of the
number $|a-b|$ in that form ($j$ is an integer).
Note that $|a|_p$ can be considered as an (ultrametric) norm of the
number $a$.

The integral over all $p$-adic numbers is defined as \cite{taibleson_75}
\begin{equation}
  \int_{p} T(a) {\rm d}a = \frac{1}{p^{K}}\sum_{a=0}^{p^{K-1}}T(a),
\end{equation}
where the sum goes over whole set of $p$-adic numbers. With so
defined integral we have the Fourier transform of the function $T(a)$
in the form:
\begin{equation}
\hat{T}(m)=\frac{1}{p^{K}}\sum_{a=0}^{p^{K}-1}e^{2\pi {\rm i}\,ma}T(a).
\end{equation}
$m$ is an element of the reciprocal space and it is in the form
$ m=\sum_{i=1}^{K} m_{i}p^{i-K}$, $0\leq m_{i}\leq p-1$. If the
function $T(a)$ depends only on the ultrametric norm, i. e.
$T(a)=f(\pam{a})$, and if $\pam{m}=p^{k}$, we get (for $k=1,\ldots,K$):
\begin{equation}
\begin{split}
\hat{T}_{k}\equiv\hat{T}(m) & = \tilde{T} + \sum_{l=k}^{K-1}
T_{l}p^{K-l-1}(p-1) - T_{k-1}p^{K-k}\\ 
\hat{T}_{0}\equiv\hat{T}(0) & = 
\tilde{T} + \sum_{l=0}^{K-1} T_{l}p^{K-l-1}(p-1)\quad.
\end{split}
\end{equation}
In the previous formula we defined,
$\tilde{T}=T(0)$, and $T_{l}=T(a)$ when $\pam{a}=p^{-l}$. On the
other side for $\pam{a}=p^{-l}$ we get the inversion of the
Fourier transform in the form (for $l=0,\ldots,K-1$):
\begin{equation}
\begin{split}
T_{l}\equiv T(a)& = p^{-K}\hat{T}_{0} +
\sum_{m=1}^{l}\hat{T}_{m}p^{m-K-1}(p-1) -
\hat{T}_{l+1}p^{l-K}\\ 
\tilde{T}\equiv T(0) & =  p^{-K}\hat{T}_{0} +
 \sum_{m=1}^{K}\hat{T}_{m}p^{m-K-1}(p-1)\quad.
\end{split}
\end{equation}

\end{document}